\documentclass[prl,twocolumn,showpacs,epsfig,preprintnumbers,amsmath,amssymb]{revtex4}

\usepackage{epsfig,subfigure}
\DeclareGraphicsExtensions{.jpg,.pdf,.eps}

\def \bea{\begin{eqnarray}}
\def \eea{\end{eqnarray}}

\begin{document}

\title{Extended time-travelling objects in Misner space}
\author{Dana Levanony and Amos Ori}
\affiliation{Department of Physics, Technion, Haifa 32000, Israel}
\date{\today}
\begin{abstract}

Misner space is a two-dimensional (2D) locally-flat spacetime
which elegantly demonstrates the emergence of closed timelike
curves from causally well-behaved initial conditions. Here we
explore the motion of rigid extended objects in this time-machine
spacetime. This kind of 2D time-travel is found to be risky due to
inevitable self-collisions (i.e. collisions of the object with
itself). However, in a straightforward four-dimensional
generalization of Misner space (a physically more relevant
spacetime obviously), we find a wide range of "safe" time-travel
orbits free of any self-collisions.

\end{abstract}

\maketitle

\section{Introduction}

About  40 years ago, Misner \cite{misner_space} introduced an
amazing vacuum solution of the Einstein equations, known as the
{\it Misner space}. This is a  two-dimensional (2D) spacetime
which describes the formation of closed timelike curves (CTCs)
from causally well-behaved initial conditions. The solution
evolves from an initial spacelike hypersurface $T=const<0$, in a
causally well-behaved manner, up to a "moment" (actually a null
hypersurface) denoted $T=0$. Subsequently, the spacetime extends
smoothly to the domain $T>0$, in which all points rest on CTCs. It
thus nicely demonstrates the phenomenon of smooth formation of
CTCs from causally well-behaved initial conditions. This solution
may straightforwardly be extended to any $d>2$ dimensions.

Quite remarkably, Misner space is actually {\it flat}.  It may be
obtained from the Minkowski spacetime by a certain cut-and-paste
operation, in a manner resembling the construction of a cone by
folding the flat Euclidean plane.

One of the outstanding open questions in spacetime physics is that
of CTCs formation: Do the laws of nature permit the creation of
CTCs from physically and causally well-behaved initial conditions?
As it stands, Misner space falls short of providing a compelling
"realistic" physical example ---mainly due to its
topologically-nontrivial ($S^1\times \mathbb{R}^{d-1}$) character
(which apparently makes it incompatible with the asymptotics of
our realistic spacetime). Several other interesting examples of
time-machine spacetimes were introduced  previously
\cite{Classic_models}, but all of them suffer from some severe
physical problems. Nevertheless, it was recently demonstrated that
a certain non-flat generalization of Misner space (based on the
"pseudo-Schwarzschild" metric rather than Minkowski) may be used
to construct a more feasible time-machine model
 \cite{Amos_Time_Machine_2007}:
Namely, an asymptotically-flat and topologically-trivial
four-dimensional (4D) spacetime which satisfies all the energy
conditions, and which smoothly develops CTCs   at a certain stage.

Beside these constructional issues, one may be concerned about
other unusual physical phenomena which may take place in a
time-machine spacetime. In particular, several authors
\cite{stability} investigated the stability of classical and
quantum fields in certain time-machine spacetimes. There are
obvious indications for linear instabilities of various kinds in
the neighborhood of the Chronology horizon \cite{stability}. It is
still unclear, however, what will be the outcome of these
instabilities in the full nonlinear context.

In this work we introduce an additional probe for the nature of
physical processes on a time-machine background: We consider the
motion of physical objects of finite size, and examine whether
such objects can penetrate (and traverse) the region of CTCs,
without being destroyed or damaged by self-collisions. For
simplicity we shall consider here the Misner space (in two or more
dimensions). Since this spacetime is flat, no tidal forces will
act on the object, which may therefore be considered as {\it
rigid}. Yet the nontrivial identifications may result in self
collisions---e.g. a "head-tail" collision of the object's two
edges.

A brief look reveals that such self-collisions certainly occur for
some orbits, but the more interesting question is whether it is
possible to choose orbits which avoid these collisions. We shall
show that it is fairly easy to avoid self-collisions up to $T=0$.
However, in the 2D case, collisions are found to be inevitable
once the object has crossed into $T>0$.

Nevertheless, we shall demonstrate here that for any $d>2$ a
collision-free motion is possible, throughout the region of CTCs,
for a wide, nonzero-measure, range of orbits. This includes the
case of most obvious physical relevance, namely that of a
three-dimensional rigid object moving in 4D Misner space.

We note that the motion of extended objects has been analyzed
previously by several authors, mostly during the 1990s, in the
context of the "billiard-ball" problem \cite{Thorne
billiard,Mensky and novikov}. To the best of our knowledge,
however, these investigations were restricted to the
wormhole-based time-machine spacetime \cite{Morris}. We are not
aware of extensions of the "billiard-ball" analyses to the
Misner-space background---or to any other background spacetime
which similarly satisfies the energy conditions \footnote{We also
note in this regard that the structure of the Cauchy/chronology
horizon in Misner space is remarkably different from that of the
wormhole-based model.}. Note also, that Misner space is especially
convenient due to its flatness, which implies vanishing tidal
forces and hence conceptually simplifies the notion of "rigid
extended object".

In section II we describe the basic structure of Misner space and
analyze its geodesics. Section III is devoted to analyzing an
extended object in a 2D Misner space, whereas in section IV we
extend Misner space to  three dimensions and analyze the object's
motion in this extended model. Section V treats the
four-dimensional case, and in section VI we briefly discuss our
results.

Throughout the paper we use relativistic units in which $c=1$.

\section{Misner space}
Misner space \cite{misner_space} is a 2D spacetime with the metric
 \bea ds^2=-2 dTd\psi-Td\psi^2, \label{2D_Misner_Metric} \eea
where $-\infty <T<\infty$ but the coordinate $\psi$ is periodic,
that is, each $\psi$ is identified with $\psi+\psi_0$ for a
certain parameter $\psi_0>0$. Since $det(g)=-1$  the metric
(\ref{2D_Misner_Metric}) is perfectly regular everywhere and in
particularly at $T=0$.

The curves $T = const$ are all closed due to the periodicity of
$\psi$. Whereas the $T<0$ curves are spacelike, the $T>0$ curves
are timelike. It then follows that all points at $T>0$ rest on
CTCs but  those at $T<0$ do not. The curve $T=0$ is null, and it
serves as the \textit{chronology horizon} (i.e. the hypersurface
separating the causal and non-causal parts of spacetime).

Any hypersurface $T=const\equiv T_0<0$ is spacelike and can be
chosen as an initial hypersurface over which initial data (for
both geometry and physical fields) may be specified. The
hypersurface $T=0$ is a Cauchy horizon  for any such initial
hypersurface $T=T_0<0$. The Cauchy evolution of the latter
unambiguously yields the portion $T_0<T<0$ of Misner space.
Assuming that the evolution beyond the Cauchy horizon proceeds in
an analytic manner, we recover the region $T>0$ as well, and CTCs
appear. Hence Misner space satisfactorily describes the formation
of CTCs from rather conventional (though topologically
non-trivial) initial conditions.

The metric (\ref{2D_Misner_Metric}) is flat, so in a local sense
it is equivalent to 2D Minkowski. However, in a global sense it is
drastically different from Minkowski due to the identification of
$\psi$. The universal covering of Misner space is obtained by
unfolding the coordinate $\psi$, namely by setting
-$\infty<\psi<\infty$. In this covering space the portions $T<0$
and $T>0$ correspond to regions I and II of Minkowski,
respectively, as shown in Fig.~\ref{fig_minkowski}. There are
infinite number of Misner copies in these two regions of
Minkowski.

We begin by presenting the Misner process --- the procedure which
transforms the Minkowski spacetime into Misner. To this end we
introduce an intermediate, Rindler-like, coordinate $z$ which will
be useful in later analysis. Following the Misner process we
derive the geodesics in the Misner coordinates and also discuss
their relation to the Rindler-like coordinate $z$.

\subsection{Coordinate transformation}\label{Sec:CT}

\begin{figure}[h]
\subfigure[]{\includegraphics[scale=0.5]{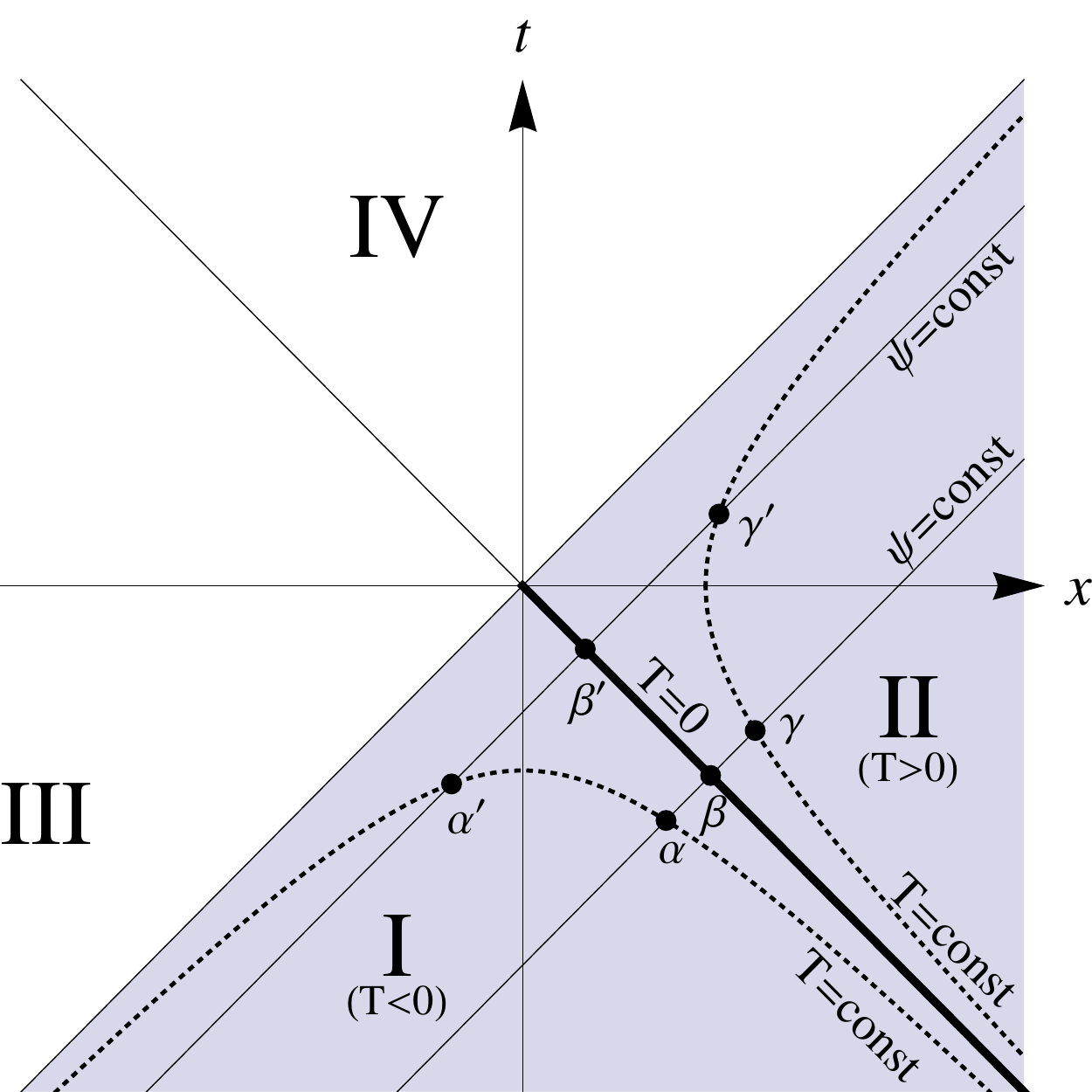}\label{fig mink
TPsi}}\ \hspace{1.2cm}
\subfigure[]{\includegraphics[scale=0.5]{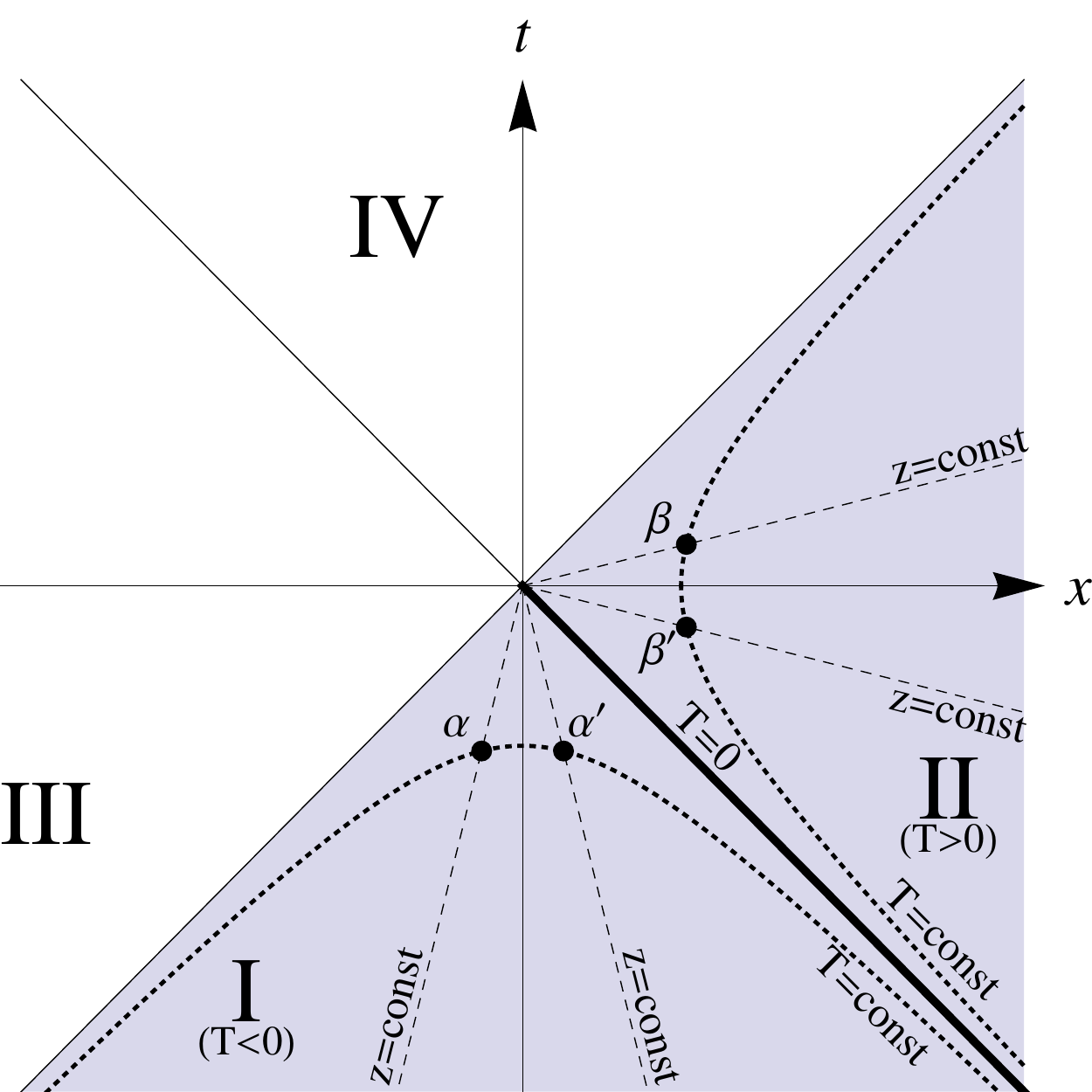}\label{fig mink
zT}} \caption{Misner space and its universal covering presented in
Minkowski coordinates ($t,x$). In both figures (a,b), the
universal covering  (the portion $t<x$, which corresponds to
$-\infty<\psi<\infty$) is the gray region, consisting of the two
Minkowski quadrants I and II. In the Misner space (compactified
$\psi$), the two diagonal lines denoted "$\psi=const$" in
Fig.\ref{fig mink TPsi} are identified. Alternatively, Misner's
identification may be implemented by identifying two $z=const$
lines. This is demonstrated in Fig.~\ref{fig mink zT}, which
displays a pair of such identified $z=const$ lines in each of the
quadrants I, II. In both figures identification points are marked
by the same Greek letter.} \label{fig_minkowski}
\end{figure}

We start with the 2D Minkowski metric \bea ds^2=-dt^2+dx^2.\eea
Misner's covering space occupies only the portion $x<t$ of
Minkowski, namely  the gray regions I and II in
Fig.~\ref{fig_minkowski}. We first elaborate on region I. We
consider the coordinate transformation \bea
x&=&-2\sqrt{-T} sinh\left(\frac{z}{2}\right),\nonumber\\
t&=&-2\sqrt{-T} cosh\left(\frac{z}{2}\right),\label{X t T z} \eea
where $T<0$ and $-\infty<z<\infty$. This leads to the metric  \bea
ds^2=\frac{dT^2}{T}-Tdz^2.\label{metric T z}\eea We now introduce
the coordinate $\psi$ by \bea \psi=z-ln|T|.\label{z psi}\eea
Transforming the line element (\ref{metric T z}) from $z$ to
$\psi$, one obtains Misner's metric (\ref{2D_Misner_Metric}).

The analytic extension of the metric (\ref{2D_Misner_Metric}) from
$T<0$ (region I) to $T>0$ (region II) is straightforward. However,
we note that the transformation (\ref{X t T z}) only applies to
region I. In order to directly transform region II  from ($t,x$)
to ($T,z$), we must modify the transformation into \bea
x&=&2\sqrt{T} cosh\left(\frac{z}{2}\right),\nonumber\\
t&=&2\sqrt{T} sinh\left(\frac{z}{2}\right).\label{X t T z TL0}
\eea It will result in the metric (\ref{metric T z}) as before.
[The transformation (\ref{z psi}) defining $\psi$ applies to $T>0$
without any modification, and yields the metric
(\ref{2D_Misner_Metric}) in region II as well.]

Note that the lines $T=const$ are spacelike in region I and
timelike in region II, and the lines $z=const$ are timelike in I
and spacelike in II. On the other hand, the lines $\psi=const$ are
everywhere null.

So far we have constructed Misner's metric
(\ref{2D_Misner_Metric}) on the (topologically-trivial)
"half-Minkowski" manifold, namely the union of regions I and II in
Fig.~\ref{fig_minkowski}. In the next stage, we choose a parameter
$\psi_0>0$ and fold the $\psi$ coordinate by identifying $\psi$
with $\psi+\psi_0$ (at the same $T$). The coordinate $T$ still
takes the entire range $-\infty<T<\infty$. A pair of such
identified constant-$\psi$ lines, embedded in the half-Minkowski
covering space, is shown in Fig.~\ref{fig mink TPsi}. On these two
lines we marked identified points by the same Greek letter. These
identified pairs of points all lie on constant-$T$ lines.

 It will be useful to note at this stage that identifying
the $\psi$ coordinate at the same $T$ value is equivalent to the
identification of the $z$ coordinate at the same $T$ [see
Eq.~(\ref{z psi})]. Lines of constant $z$ are presented in
Fig.~\ref{fig mink zT}, along with the constant-$T$ lines. Again,
we marked identified points by the same Greek letter.

Altogether the transformation from Minkowski to Misner is: \bea
t=Te^{\frac{\psi}{2}}-e^{-\frac{\psi}{2}},\nonumber\\
x=Te^{\frac{\psi}{2}}+e^{-\frac{\psi}{2}}, \label{X and t function
of} \eea and the inverse transformation is: \bea
\psi&=&-2\ln\left(\frac{x-t}{2}\right),\label{T Psi of X t1}\\
T&=&\frac{x^2-t^2}{4}.\label{T Psi of X t2} \eea These relations
hold at both regions I and II.

As was mentioned above, Misner's identification can be manifested
by identifying two lines of constant $z$ (at the same $T$, and
with $z$ values separated by $\psi_{0}$). The velocity $dx/dt$ is
fixed along each such line of constant $z$ {[}cf. Eqs. (\ref{X t T
z}) or (\ref{X t T z TL0}){]}, therefore the relative velocity
between a pair of identified $z=const$ lines is well defined. A
straightforward calculation reveals that this relative velocity is
$u=Tanh(\psi_{0}/2)$. Thus, Misner's {}``folding'' process may be
viewed as identification under the action of a boost with velocity
$u$.

Let (p,q) be a pair of points in the half-Minkowski covering
space, and let (p',q') be their images under a certain (generic)
boost. Since in 2D Minkowski spacetime any two boosts are
commutative, it immediately follows that p' and q' are identified
(under Misner's folding) if and only if p and q are identified. It
thus follows that the 2D Misner space inherits the boost
invariance of Minkowski.

\subsection{Geodesics}\label{sec:geodesics}
Our main interest in this work is the motion of a rigid object in
Misner space. To simplify the analysis, we shall employ the above
mentioned boost symmetry and choose a Lorentz frame in which the
object is at rest [namely, $x(t)=const$]. We start here by
analyzing the properties of a single such geodesic.

It is convenient to express the geodesics using their
corresponding function $T(\psi)$ \footnote{Since the $\psi$
coordinate is null, it increases monotonically along any timelike
geodesic and is suitable to use as a parameter.}. A single static
geodesic satisfies (in the covering space) $x=const\equiv x_0$,
which by virtue of Eq.~(\ref{X and t function of}) yields \bea
T(\psi)=-e^{-\psi}+x_0e^{-\frac{\psi}{2}} .\label{geodesic Tpsi}
\eea

Consider the propagation of such an $x=x_0$  geodesic from some
$T<0$ toward $T=0$. The relation (\ref{geodesic Tpsi}) makes it
clear that there are two different classes of such geodesics:
Those with $x_0<0$ only approach $T=0$ at $\psi \to \infty$. On
the other hand, those with $x_0>0$ will all reach $T=0$ at a
\textit{finite} $\psi$, and continue their journey in the region
$T>0$ \footnote{By using a different coordinate transformation one
can extend region I into region III \cite{misner_space,
Hawking_Book} instead of II. In these alternative coordinates, the
two classes of geodesics ($x_0>0$ and $x_0<0$) interchange their
role.}. Since our primary objective is the motion of extended
objects into the region of CTCs ($T>0$), throughout the rest of
the paper we shall restrict our attention to the second class,
namely the $x_0>0$ geodesics.

Consider now the behavior of those $x_0>0$ geodesics at $T>0$. For
each of these geodesics the function $T(\psi)$ will reach its
maximum at its intersection point with $t=0$. This behavior is
demonstrated in Fig.~\ref{fig geo Tpsi}, which displays two
different $x=x_0>0$ geodesics, as well as the line $t = 0$. This
property can be easily deduced by finding the $(T,\psi)$
coordinates at the maximum point of the relation (\ref{geodesic
Tpsi}), and using Eq.~(\ref{X and t function of}) in order to
obtain the corresponding $t$ value.

These static geodesics exhibit a simple symmetry when displayed in
the $(T,z)$ coordinates. From Eq.~(\ref{X t T z TL0}) we observe
that in the $T>0$ region, the relation $x=x_0$ yields \bea
T(z)=\left(\frac{x_0}{2 cosh(z/2)}\right)^2. \label{Geo_Tz}\eea
This function is symmetric about $z=0$, hence the maximum of $T$
is attained at $z=0$. This is illustrated in Fig.~\ref{fig geo
Tz}, which displays a single $x=x_0$ geodesic and the line $t=0$
in $(T,z)$ coordinates. From Eq.~(\ref{X t T z TL0}) it is also
clear that $t=0$ coincides with $z=0$, demonstrating again that
the geodesics $x=x_0$ reach their maximal $T$ value at a point
where $t$ vanishes.

\begin{figure}[h]
\subfigure[]{\includegraphics[scale=0.8]{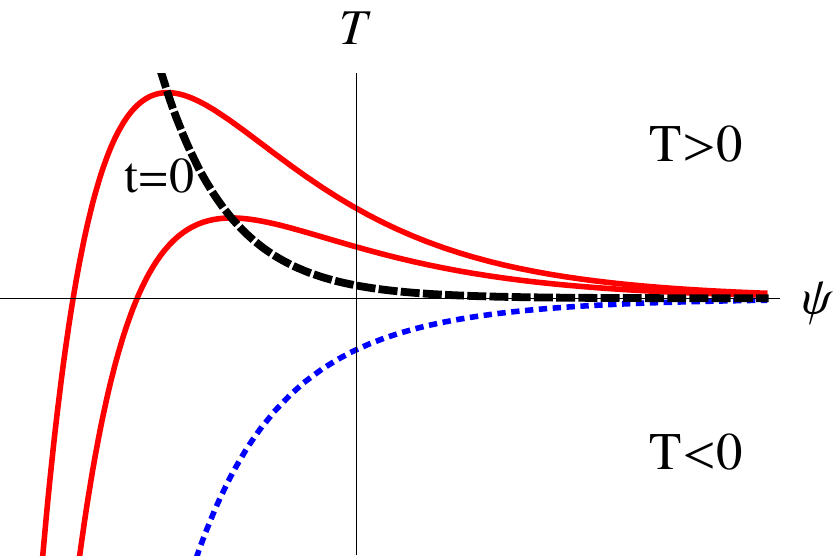}\label{fig geo
Tpsi}}\ \subfigure[]{\includegraphics[scale=0.8]{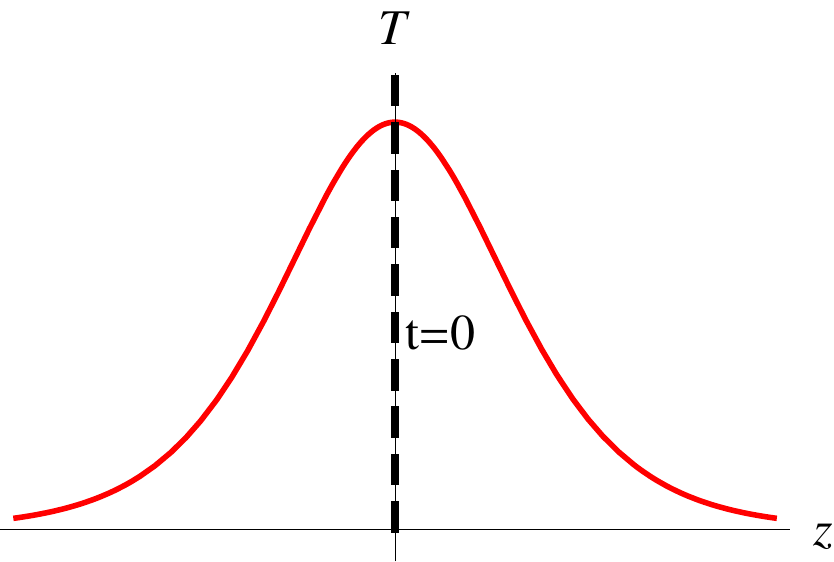}\label{fig
geo Tz}} \caption{Constant-$x$ geodesics plotted in Misner
coordinates ($T,\psi$) and in ($T,z$) coordinates. The dashed
black curve represents the $t=0$ line, which in fact coincides
with $z=0$. Fig.~\ref{fig geo Tpsi} shows three geodesics: The
dotted (blue) curve is an $x_0<0$ geodesic. The two solid (red)
curves represent $x_0>0$ geodesics. These geodesics attain their
maximal $T$ values at their intersection with $t=0$. Fig.~\ref{fig
geo Tz} demonstrates the symmetry of a single ($x_0>0$) geodesic
around $z=0$.} \label{fig_geodesic}
\end{figure}

\section{Rod motion: the two dimensional case}

Consider now a rigid extended object, a "rod", which moves freely
in 2D Misner space. The rod may be considered as a one-parameter
family of \textit{rod's points}. Presumably no external forces are
present,  and since Misner space is flat, the tidal force vanishes
as well, so all rod's points are expected to move on geodesics (as
long as self-collisions have not occurred). The rod's motion in
spacetime is thus described by a congruence of timelike geodesics.
Rigidity implies that these geodesics are all parallel (in $x$-$t$
coordinates). However, the identification of $\psi$ may lead to a
collision of two rod's points. Furthermore, at $T>0$ a rod point
may even collide with itself (owing to the presence of CTCs). Our
main objective is to investigate whether such collisions may be
prevented.

Exploiting the boost-invariance of Misner space, we choose a
Lorentz frame in which the rod is at rest (in the corresponding
Minkowskian universal-covering space), so all geodesics in the
rod's congruence satisfy $x=const\equiv x_0$. Each rod's point is
thus characterized by its $x_0$ value.

The rod presumably starts its journey at the pre-CTCs region
$T<0$, and moves towards the CTCs region $T>0$. We shall first
consider the journey toward the chronology horizon, namely the
domain $T<0$.

\begin{figure}[h]
\subfigure[]{\includegraphics[scale=0.35]{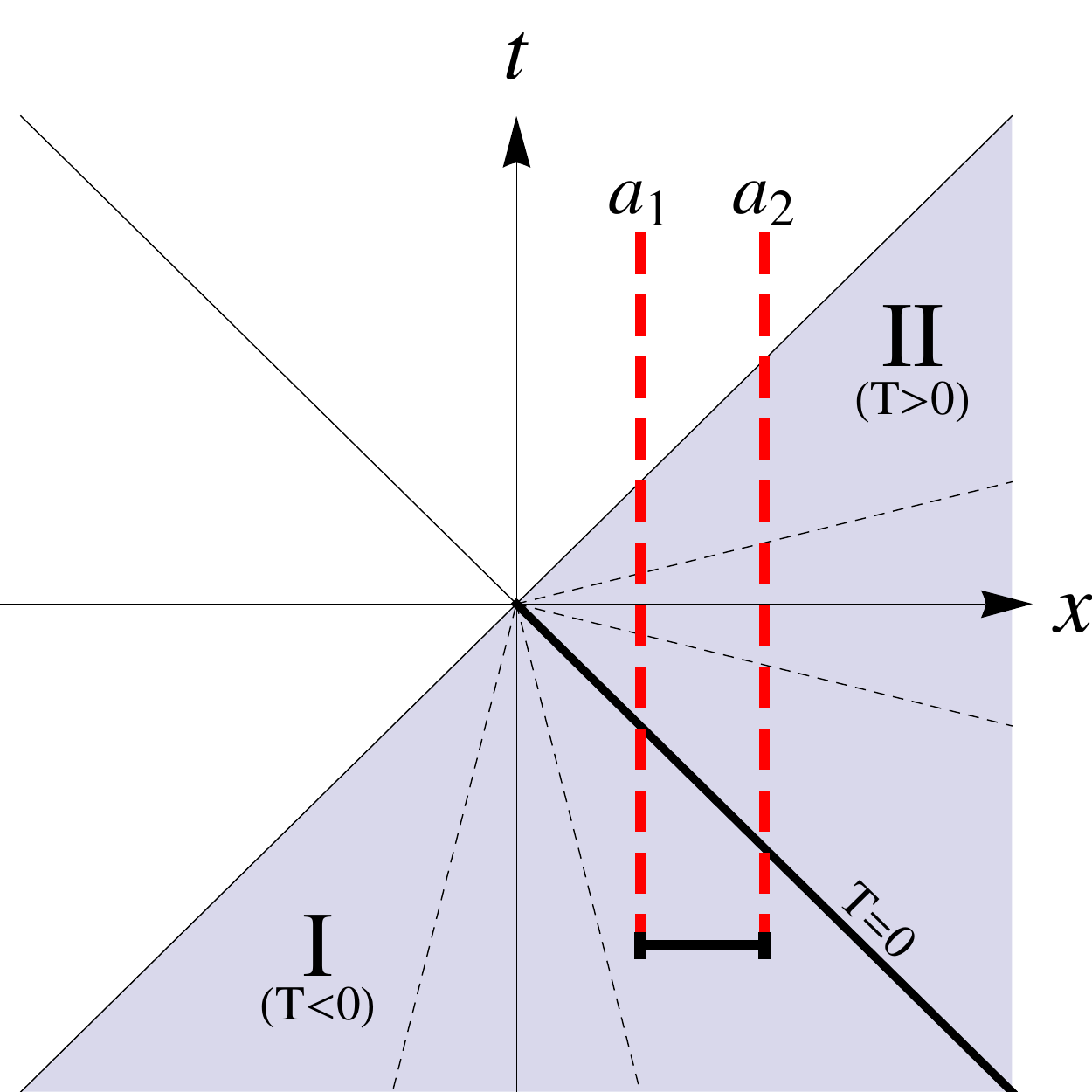}\label{fig geo 4D
Mink}}\ \subfigure[]{\includegraphics[scale=0.6]{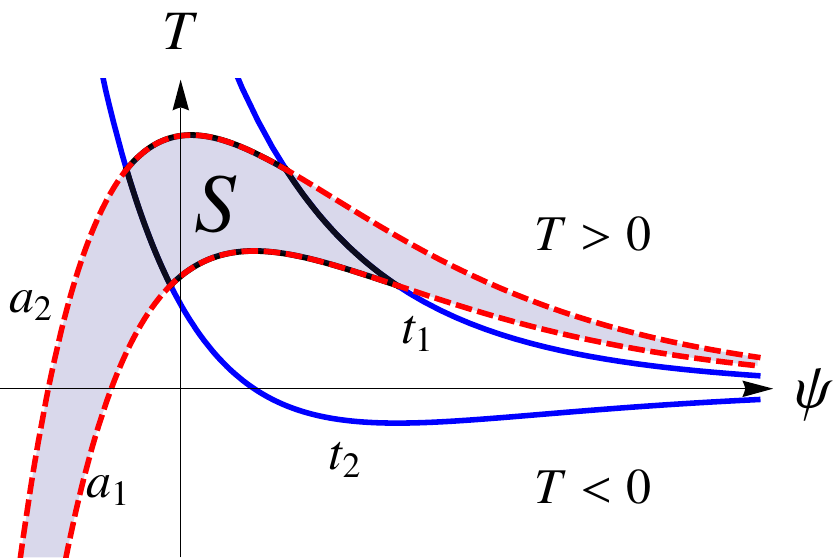}\label{fig
geo 4D Tpsi}}\
\subfigure[]{\includegraphics[scale=0.6]{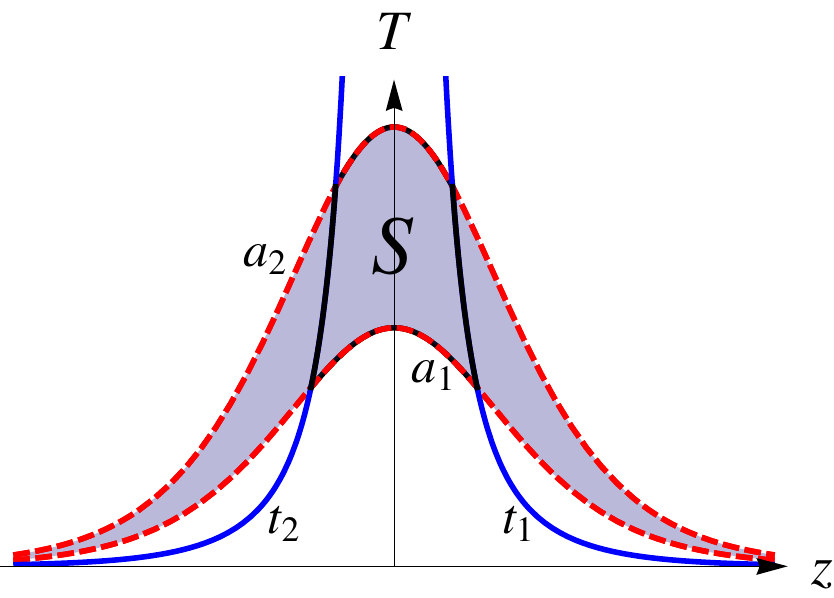}\label{fig geo 4D
Tz}} \caption{A  plot of the rod's two edge geodesics $x=a_{1,2}$,
embedded in the universal covering space. These geodesics are
represented by dashed (red) lines --- in Minkowski coordinates
($t,x$), in Misner coordinates ($T,\psi$), and in ($T,z$)
coordinates, in Figs. 3(a,b,c) respectively. In Fig. 3(a) the
short horizontal bold solid (black) line represents the rod, and
the two vertical lines are the edge geodesics. In Figs. 3(b,c) the
two solid (blue) curves represent curves of constant $t$. In the
2D case, the spacetime region occupied by the rod is the entire
gray strip. In the $d \geq 3$ case (with $v \neq 0$, as discussed
below), at a given $y=const$ hypersurface the rod occupies the
domain $S_y$, namely the quadrangle-like region denoted "S".
}\label{fig_geodesic3}
\end{figure}

We denote the rod's two edges by $x_0=a_1$ and $x_0=a_2$, assuming
$0<a_1<a_2$. Figs.~\ref{fig_geodesic3}(a-c) display the orbits of
the two edge geodesics (in the universal covering space) by dashed
(red) lines, in $t$-$x$, $T$-$\psi$, and $T$-$z$ coordinates,
using Eqs.~(\ref{geodesic Tpsi}) and (\ref{Geo_Tz}). The rod thus
occupies the region between the two red lines, marked by gray in
Figs.~\ref{fig geo 4D Tpsi} and \ref{fig geo 4D Tz}. Evidently, a
necessary and sufficient condition for a safe journey up to the
chronology horizon will be that at any slice $T=constant \leq 0$,
the $\psi$-difference between the two edges will be $<\psi_0$.

The function $T(\psi)$ of Eq.~(\ref{geodesic Tpsi}) is monotonic
throughout the region $T \leq 0$ and can thus be inverted: \bea
\psi(T)=2\ln2-2\ln[x_0+(x_0^2-4T)^{1/2}].\label{inverted geodesic}
\eea At constant $T$, the $\psi$-difference between the two edges
will be \bea \Delta \psi(T)\equiv |\psi_2(T)-\psi_1(T)|=2\ln
\left(\frac{a_2+\sqrt{a_2^2-4T}}{a_1+\sqrt{a_1^2-4T}}\right).\label{delta
psi1} \eea A collision-free motion will occur  if $\Delta \psi(T)<
\psi_0$ for any $T \leq 0$. Since the right-hand side of
Eq.~(\ref{delta psi1}) is a monotonically increasing function of
$T$, it will reach its maximum (in the domain $T\leq0$ presently
under consideration) at $T=0$. This determines the criterion for a
collision-free motion of the rod up to the chronology horizon:
$\psi_0>2\ln(a_2/a_1)$. This criterion may be reformulated as a
condition on $a_1$: \bea a_1>\frac{L_x}{e^{\psi_0/2}-1},
\label{Rod_passing_Criterion2}\eea where $L_x \equiv a_2-a_1$ is
the rod's length.

We turn now to consider the rod's motion in the CTCs region $T>0$.
As is evident from Eq.~(\ref{geodesic Tpsi}), the $x_0>0$
geodesics reach a (positive) maximal  value of $T$, then $T(\psi)$
decreases monotonically until it vanishes as $\psi \to \infty$.
This behavior is demonstrated by the two red lines in
Fig.~\ref{fig geo 4D Tpsi}. We denote this (geodesic-dependent)
maximal $T$ value by $T_{max}$. Any constant-$T$ line in the range
$0<T<T_{max}$ intersects the geodesic {\it twice}, at two
different $\psi$ values. For a given geodesic, we denote the
$\psi$-difference between these two intersection points by $\delta
\psi (T)$. This $\delta \psi (T)$ diverges at the limit $T\to
0_+$. A self-collision occurs whenever $\delta \psi (T)=n \psi_0$,
where hereafter $n$ denotes a nonvanishing integer. Thus,
regardless  of the value of $\psi_0$, there will be an infinite
sequence of such self-collisions as $T\to 0_+$. This means that
any timelike geodesic will hit itself an infinite number of times
immediately after crossing $T=0$.

We conclude, that within the framework of 2D Misner space, once a
point particle crosses the chronology horizon it will inevitably
hit itself. It is obvious that a finite-length rod will be subject
to such self-collisions too.

However, our physical spacetime is four-dimensional. Can these
extra dimensions save the object from this inevitable fate of
self-collisions? The rest of this paper will be devoted to
addressing this question. We shall show that by adding one
dimension (or more) to the Misner space, the way is opened for a
collision-free journey of rigid objects. We shall first
demonstrate this in three dimensions, and then address the
straightforward extension to four (or more) dimensions.

\section{The three-dimensional case}
We shall consider now a two-dimensional rigid extended object
moving in a three-dimensional spacetime with the flat metric \bea
ds^2=-2dTd\psi-Td\psi^2+dy^2,\eea which is the straightforward
extension of the 2D Misner  metric (\ref{2D_Misner_Metric}) to
three dimensions. As before, $\psi$ is periodic with a period
$\psi_0$, and $-\infty < T,y <\infty $. Using Eq.~(\ref{X and t
function of}) again to transform ($T,\psi$) into ($t,x$), one
recovers the standard three-dimensional Minkowski line element in
the Cartesian coordinates ($t,x,y$).

As was mentioned above, Misner's identification in the $t$-$x$ (or
$T$-$\psi$) plane may be associated to a boost (with relative
velocity $u$) in the $x$ direction. Now, in $d>2$ Minkowski
spacetime two boosts commute if and only if they are co-directed.
By a straightforward extension of the discussion at the end of
Sec.~\ref{Sec:CT} we observe that $d>2$ Misner space is invariant
to boosts in the $x$ direction, but not to boosts in any other
direction.

Similar to the two-dimensional case, we assume that all object's
points (OPs) move along geodesics. In the Minkowski coordinates
these are just straight lines. The object's rigid motion in
spacetime is described by a congruence  of parallel timelike
geodesics, all sharing the same velocity vector. As before, we use
the boost invariance in the $t-x$ plane to pick a Lorentz frame in
which the object's velocity has vanishing $x$ component. We
assume, however, that the object does have a nonvanishing velocity
$v>0$ in the $y$ direction (otherwise, the previous analysis would
still hold at each $y$ separately, and self-collisions would be
inevitable at $T>0$). The OPs thus move along the geodesics \bea
x(t)= x_0 \, , \, \, \, y(t)=y_0 + v t , \label{xyMotion} \eea
where the constants of motion $x_0,y_0$ characterize the object's
individual points. For simplicity we shall consider here a
rectangular object described by \bea a_1\leq x_0 \leq a_2 \, ,
\,\,\,  b_1 \leq y_0 \leq b_2 \label{shape} \eea with $a_1 >0$
\footnote{Even if the object is not rectangular, as long as it is
compact it may be contained in such a rectangle. (The survival of
a rectangular object, obviously implies the survival of any
smaller object contained in that rectangle.)}. The object's
dimensions (in the chosen Lorentz frame) are $\ell^x=a_2-a_1$,
$\ell^y=b_2-b_1$. The {\it proper} dimensions (as measured in the
object's local rest frame) are $L^{x}= \ell^{x}$ and $L^y=\gamma
\ell^y$, where $\gamma=1/\sqrt{1-v^2}$.

A collision occurs when two distinct events (p,q) on the object's
congruence satisfy \bea
T_p=T_q,\,\,\,\psi_p=\psi_q+n\psi_0,\,\,\,y_p=y_q .
\label{collide} \eea (These two events either belong to two
different object's geodesics, or to the same geodesic but at
different proper times.)

We shall analyze the possibility to avoid self collisions ---
first at $T\leq 0$, and then at $T>0$.

\subsection{$T\leq 0$}

It is easy to see that Eq.~(\ref{Rod_passing_Criterion2}) remains
a sufficient condition for collision-avoidance at $T\leq0$: First,
any collision in 3D must involve, in particular, a collision in
the two-dimensional subspace ($T,\psi$) [as manifested by the
first two equalities in (\ref{collide})]. Also the relation
(\ref{xyMotion}) for $x(t)$ is the same as it was in the 2D case.
Thus, the analysis of the previous section still implies that if
Eq.~(\ref{Rod_passing_Criterion2}) is satisfied, collisions in the
($T,\psi$) subspace will be avoided at $T\leq0$.

\subsection{$T>0$}

Let us denote by S$_y$ the intersection of the object's congruence
(a three-parameter set) with some $y=const$ slice. S$_y$ is a
two-parameter set which may be parametrized by ($x_0,y_0$). It
occupies a nonzero-measure portion of the $y=const$ hypersurface.
We may use ($T,\psi$) as coordinates for this hypersurface---and
hence also for its subset S$_y$. Fig.~\ref{fig geo 4D Tpsi}
displays a certain $y=const$ hypersurface and its corresponding
subset S$_y$, which is the quadrangle-like domain denoted "S". The
association of an OP ($x_0,y_0$) with the corresponding ($T,\psi$)
coordinates (for a specific $y$) is done by Eqs. (\ref{T Psi of X
t1},\ref{T Psi of X t2})
--- along with Eq. (\ref{xyMotion}), which now reads $x=x_0$ and
$t=(y-y_0)/v$. The boundary of S$_y$ thus consists of the two
lines $x=a_{1,2}$ and the two lines $t=t_{1,2}$, where
$t_{1,2}\equiv (y-b_{1,2})/v$. From Eq.~(\ref{X and t function
of}), each of these boundary lines corresponds to a curve in the
($T,\psi$) plane, given by either
$T=x_{1,2}\,e^{-\psi/2}-e^{-\psi}$ or
$T=t_{1,2}\,e^{-\psi/2}+e^{-\psi}$.

For later convenience we define $\Delta t \equiv t_1-t_2$ and
$t_{cen} \equiv (t_1+t_2)/2$, such that $t_{1,2}=t_{cen} \pm
\Delta t /2$. Note that $\Delta t =\ell^y/v$ is a constant of
motion.  On the other hand $t_{cen}$ grows linearly with $y$:
$t_{cen}(y)=-(b_1+b_2)/2v+y/v$. This will allow us to replace the
variable $y$ by $t_{cen}$ in the analysis below.

For any line $T=const$ which intersects S$_{y}$, we define
$\Delta\psi_{y}(T)$ to be the span of $\psi$ along the
intersection of this line with S$_{y}$. More precisely,
$\Delta\psi_{y}(T)$ is the $\psi$-difference between the two
(furthest \footnote{For certain $y$ and $T$ values, the number of
intersection points will be four rather than two.}) intersection
points of the line $T=const$ with the boundary of S$_{y}$. We
further define $\Delta\psi_{max}(y)\equiv
\displaystyle\max_{T>0}\Delta\psi_{y}(T)$. It now immediately
follows from Eq. (\ref{collide}) that a collision may occur at a
given $y$ only if $\Delta\psi_{max}(y)\geq\psi_{0}$. That is, a
collision-free motion is guaranteed if
$\Delta\psi_{max}(y)<\psi_{0}$ for all $y$. This raises the issue
of whether $\Delta\psi_{max}(y)$ is bounded (as a function of $y$)
or not.

It will be easier to explore the dependence of $\Delta \psi_{max}$
on $t_{cen}$ than on $y$. We thus define \footnote{One can easily
verify that in the portion $T>0$ of the object's congruence
$|t_{cen}|$ is bounded by $a_2+\Delta t /2$. Here and in Eq.
(\ref{eq:DELTA_MAX}) $t_{cen}$ is confined to this range.}
\begin{equation}
\Delta\Psi\equiv\max_{t_{cen}}\Delta\psi_{max}(t_{cen}).\end{equation}
If this maximum exists (i.e. it is finite), then the condition for
collision-free motion will be simply $\psi_0>\Delta\Psi$.

\begin{figure}[h]
\includegraphics[scale=1]{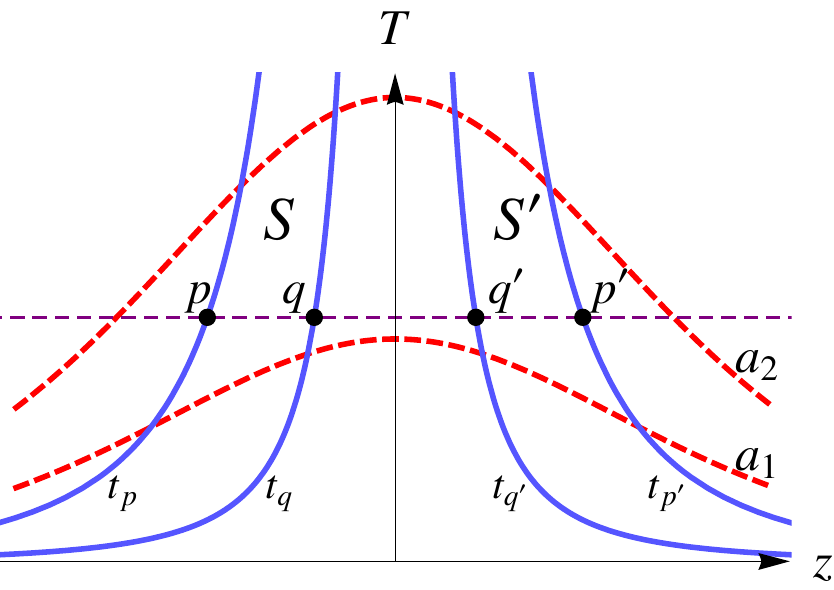}\caption{The reflection symmetry of the
geodesics and constant-$t$ lines around the $z=0$ line,
illustrated in  ($T,z$) coordinates. The dashed (red) curves
represent the geodesics of the rod's two edges, $x=a_{1,2}$. The
four solid (blue) curves represent curves of constant $t$. S and
S' are two symmetric S$_y$ regions, corresponding to two different
$y=const$ slices with the same $|t_{cen}|$.}\label{fig 2S}
\end{figure}

We shall now employ the reflection symmetry of the problem with
respect to the $z$ coordinate to show that it is sufficient to
take the maximum of $\Delta\psi_{max}(t_{cen})$ in the range
$t_{cen}<0$. To this end all we need to show is that the function
$\Delta\psi_{max}(t_{cen})$ is symmetric about $t_{cen}=0$. This
symmetry is illustrated in Fig.~\ref{fig 2S}, which displays (in
$T$-$z$ coordinates) two different, symmetric, S$_{y}$ regions,
one (denoted S) for which $t_{cen}=-t_{0}$ and the other one
(denoted S') for which $t_{cen}=t_{0}$, for some $t_{0}>0$. The
horizontal dashed (purple) line denotes a certain $T=const$ line.
The figure also shows the four $t=const$ lines which border these
two S$_{y}$ regions---as well as their four intersection points
with the $T=const$ line, denoted by p, q, q' and p'.
Correspondingly we denote the four $t$ values by $t_{p}$, $t_{q}$,
$t_{q'}$, and $t_{p'}$ respectively. Since $|t_{cen}|$ is the same
for S and S' and $\Delta t$ is fixed, one can easily verify that
$t_{p'}=-t_{p}$ and $t_{q'}=-t_{q}$. Consider now the pair of
points p and p'. They have a common $T$ but opposite $t$ values
($t_{p'}=-t_{p}$). From Eq.~(\ref{X t T z TL0}) it follows that
these two points also have opposite $z$ values, $z_{p'}=-z_{p}$.
The same argument obviously applies to the other pair of points q,
q', and we obtain $z_{q'}=-z_{q}$. Defining
$\Delta_{z}=z_{q}-z_{p}$ and $\Delta_{z}'=z_{p'}-z_{q'}$, we find
that $\Delta_{z}'=\Delta_{z}$. However, from Eq.~(\ref{z psi})  it
is obvious that for any pair of points on a given $T=const$ line,
the differences in $\psi$ and in $z$ are the same. Therefore,
$\Delta\psi_{y}(T)$ (defined above) is the same for S and S'.
Since this argument applies to \emph{any} $T=const$ line
\footnote{In the situation depicted in Fig.~\ref{fig 2S} (and
discussed in the text), the line $T=const$ intersects S at its two
constant-$t$ boundaries (and the same applies to S'). There exists
other $T=const$ lines, however, which intersect the boundary of S
at a constant-$x$ line---say, $x=a_1$. The argument presented in
the text can be easily extended to such $T=const$ lines as well,
yielding again the same $\Delta\psi_{y}(T)$ for S and S'.}, we
find that $\Delta\psi_{max}$ is also the same for these two
symmetric S$_{y}$ regions, which completes our argument. We
therefore conclude that
\begin{equation}
\Delta\Psi=\max_{t_{cen}\leq0}\Delta\psi_{max}(t_{cen}).\label{eq:DELTA_MAX}\end{equation}

Next, for any S$_{y}$ we define $\Delta\tilde{\psi}_{\max}$ to be
the maximal $\psi$-difference between \textit{all points} in
S$_{y}$. Obviously, for any $T$ (and any given S$_{y}$),
$\Delta\tilde{\psi}_{\max}\geq\Delta\psi_{y}(T)$, therefore
$\Delta\tilde{\psi}_{\max}\geq\Delta\psi_{max}$. As is obvious
from the layout of S$_{y}$ in Fig.~\ref{fig geo 4D Tpsi},
$\Delta\tilde{\psi}_{\max}$ is nothing but the $\psi$-difference
between the two {}``corners'' $(x=a_{1},t=t_{1})$ and
$(x=a_{2},t=t_{2})$ of S$_{y}$, and from Eq.~(\ref{T Psi of X t1})
it immediately follows that \begin{equation}
\Delta\tilde{\psi}_{\max}=2\ln\frac{a_{2}-t_{2}}{a_{1}-t_{1}}=2\ln\frac{a_{2}+\Delta
t/2-t_{cen}}{a_{1}-\Delta t/2-t_{cen}}.\end{equation} We now
define, in analogy with Eq. (\ref{eq:DELTA_MAX}) above,
\begin{equation}
\Delta\tilde{\Psi}=\max_{t_{cen}\leq0}\Delta\tilde{\psi}_{\max}(t_{cen}).\label{eq:DELTA_MAX_tild}\end{equation}
Obviously, $\Delta\tilde{\Psi}\geq\Delta\Psi$, hence finiteness of
$\Delta\tilde{\Psi}$ will guarantee a finite $\Delta\Psi$---and
will ensure collision-free motion for any
$\psi_0>\Delta\tilde{\Psi}$.

The maximum in Eq. (\ref{eq:DELTA_MAX_tild}) is easily calculated.
Notice that $\Delta\tilde{\psi}_{\max}$ is a
monotonically-increasing function of $t_{cen}$, therefore the
maximum is attained at $t_{cen}=0$:\begin{equation}
\Delta\tilde{\Psi}=\Delta\tilde{\psi}_{\max}(t_{cen}=0)=2\ln\frac{a_{2}+\Delta
t/2}{a_{1}-\Delta t/2}.\end{equation} Clearly, this parameter is
well-defined only if $a_{1}>\Delta t/2$, which we shall assume.

As mentioned above, a sufficient condition for a collision-free
motion is $\psi_0>\Delta\tilde{\Psi}$. It will be useful to
re-express this last inequality as a condition on $a_{1}$, once
$\psi_0$ is given. Setting $a_{2}=a_{1}+L_{x}$ one obtains the
condition
\begin{equation} a_{1}>\frac{L^{x}+\Delta
t}{e^{\psi_{0}/2}-1}+\frac{\Delta
t}{2}.\label{result}\end{equation} Note that this inequality
automatically ensures that $a_{1}>\Delta t/2$ (which was assumed
above).

This condition on $a_{1}$ was designed so as to avoid collisions
throughout the region $T>0$. However, it is definitely stronger
than the inequality (\ref{Rod_passing_Criterion2}) which ensured
collision-free motion at $T\leq0$. We therefore conclude that the
constraint (\ref{result}) is \emph{a sufficient condition for
avoiding collisions throughout the entire Misner space}. Stated in
other words: Given the spacetime's identification parameter
$\psi_{0}$, the object's dimensions $L^{x},L^{y}$, and its
velocity $v>0$ (and hence also $\Delta t=L^{y}/\gamma v$), it is
always possible to avoid collisions by placing the object at
sufficiently large $x$ values---namely, by increasing $a_{1}$.

\section{The four-dimensional case}

We turn now to consider the more realistic case, the {\it
four-dimensional}  Misner space with the metric \bea
ds^2=-2dTd\psi-Td\psi^2+dy^2+dZ^2 \eea (with periodic $\psi$ as
before, and $-\infty <T,y,Z< \infty $). The object again has a
velocity $v>0$ in a direction perpendicular to $x$, and without
loss of generality we take it to be in the $y$ direction. The OPs
thus move on parallel geodesics satisfying Eq. (\ref{xyMotion}) as
well as $Z(t)=Z_0$. The object is now assumed to be a
three-dimensional rectangular box described by Eq. (\ref{shape})
along with $c_1 \leq Z_0 \leq c_2$.

One can easily verify that since there is no motion  in the $Z$
direction (unlike in $y$), the addition of the $Z$ dimension does
not affect the above analysis in any way (that is, the analysis of
the previous section still applies at any $Z=const$ slice).
Equation (\ref{result}) thus remains a sufficient condition for a
collision-free motion.

\section{Discussion}
We conclude that self-collisions indeed constitute a real threat
for time travels, but at the same time they do not pose an
impenetrable barrier: In the four-dimensional Misner space (like
in any of its $d\geq 3$ counterparts), there exists a wide range
in the space of possible orbits for which self-collisions are
avoided---as demonstrated in Eq.~(\ref{result}). However, this
requires the object to have a sufficient velocity in a direction
perpendicular to the one underlying the Misner identification.

As was discussed above, the Misner space itself admits a
non-standard  topology ($\psi$ is closed), which restricts the
physical relevance of this specific flat geometry. However,
curved-spacetime generalizations of 4D Misner space (e.g. the
compactified "pseudo-Schwarzschild" geometry) may serve as a core
for more acceptable time-machine spacetimes, which are
topologically-trivial and asymptotically-flat
\cite{Amos_Time_Machine_2007}. It will be interesting to
investigate the motion of extended objects into and throughout the
CTCs region of such non-flat time-machine spacetimes as well.

This research was supported in part by the Israel Science
Foundation (grant no. 1346/07).

\end{document}